\documentclass[a4paper,12pt]{article}
\usepackage{epsfig}
\usepackage[dvips,usenames]{color}
\usepackage{graphicx}
\def\pslash{\rlap{\hspace{0.02cm}/}{p}}

\newlength{\dinwidth}
\newlength{\dinmargin}
\begin{document}
\title{Pair production of Higgs bosons associated with $Z$ boson in the left-right twin Higgs model at the ILC}
\bigskip
\author{Yao-Bei Liu$^{1}$, Xue-Lei Wang$^{2}$\\
{\small 1: Henan Institute of Science and Technology, Xinxiang
453003, P.R.China}
\thanks{E-mail:liuyaobei@sina.com}\\
{\small 2: College of Physics and Information Engineering,} \\
{\small Henan Normal University, Xinxiang 453007, P.R.China}\\ }
\maketitle
\begin{abstract}
\indent The left-right twin Higgs(LRTH) model predicts the existence
of three additional Higgs bosons: one neutral Higgs $\phi^{0}$ and a
pair of charged Higgs bosons $\phi^{\pm}$. In this paper, we studied
the production of a pair of charged and neutral Higgs bosons
associated with standard model gauge boson $Z$ at the ILC, i.e.,
$e^{+}e^{-}\rightarrow Z\phi^{+}\phi^{-}$ and $e^{+}e^{-}\rightarrow
Z\phi^{0}\phi^{0}$. We calculate the production rate and present the
distributions of the various observables, such as, the distributions
of the energy and the transverse momenta of final $Z$-boson and
charged Higgs boson $\phi^{-}$, the differential cross section of
the invariant mass of charged Higgs bosons pair, the distribution of
the angle between charged Higgs bosons pair and the production angle
distributions of $Z$-boson and charged Higgs boson $\phi^{-}$. Our
numerical results show that, for the process $e^{+}e^{-}\rightarrow
Z\phi^{+}\phi^{-}$, the production rates are at the level of
$10^{-1} fb$ with reasonable parameter values. For the process of
$e^{+}e^{-}\rightarrow Z\phi^{0}\phi^{0}$, we find that the
production cross section are smaller than $6\times 10^{-3} fb$ in
most of parameter space. However, the resonance production cross
section can be significantly enhanced.
\end{abstract}
PACS number(s): 12.60.Fr, 13.66.Hk, 14.70.Hp\\
Keywords: Left-right twin higgs model, charged Higgs boson, ILC.
\newpage
\noindent{\bf I. Introduction}\\
\indent One interesting approach to the hierarchy problem, first
proposed in \cite{hierarchy-1,hierarchy-2}, is that the Higgs mass
parameter is protected because the Higgs is the pseudo-Goldstone
boson of an approximate global symmetry. In the last few years
several interesting realizations of this idea based on the little
Higgs mechanism have been constructed \cite{little-1,little-2}.
These theories stabilize the weak scale up to be above a few TeV.
Many alternative new physics theories, such as supersymmetry,
topcolor, and little Higgs, predict the existence of new scalar or
pseudo-scalor particles. These new particles may have cross sections
and branching fractions that differ from those of the SM Higgs
boson. Thus, the discovery of the new scalars at the future high
energy colliders might shed some light on the new
physics models.\\
\indent Recently, the twin Higgs mechanism has been proposed as an
interesting solution to the little hierarchy problem
\cite{ly-1,ly-2,ly-3}. The SM Higgs emerges as a pseudo-Goldstone
boson once a global symmetry is spontaneously broken, which is
similar to what happens in the little Higgs models \cite{little-1}.
Gauge and Yukawa interactions that explicitly break the global
symmetry give mass to the Higgs. Once an additional discrete
symmetry is imposed, the leading quadratic divergent term respects
the global symmetry, thus does not contribute to the Higgs mass. The
twin Higgs mechanism can be implemented in left-right models with
the discrete symmetry being identified with left-right symmetry
\cite{ly-2}. The left-right twin Higgs(LRTH) model contains
$U(4)_{1}\times U(4)_{2}$ global symmetry as well as
$SU(2)_{L}\times SU(2)_{R}\times U(1)_{B-L}$ gauge symmetry. In the
LRTH model, pair of vector-like heavy top quarks play a key role at
triggering electroweak symmetry breaking just as that of the little
Higgs theories. Besides, the other Higgs particles acquire large
masses not only at quantum level but also at tree level. The
phenomenology of the LRTH model are widely discussed in literature
\cite{Hock,dong}, and constraints on LRTH model parameters are
studied in \cite{loop}. The LRTH model is also expected to give new
significant signatures in future high energy colliders and studied
in references \cite{liu}, due to the new particles which are
predicted by this model. Also the pair production of the charged and
neutral Higgs bosons at the ILC and LHC in the framework of the LRTH
model are studied
in \cite{liu1, yue}.\\
\indent The hunt for the Higgs boson and the elucidation of the
mechanism of symmetry
 breaking is one of the most important goals for present and future
 high energy collider experiments. The most precise measurements
 will be performed in the clean environment of the future $e^{+}e^{-}$ linear
 colliders, with a center of mass(c.m.) energy in the range of 500 to 1600$GeV$,
 as in the case of the International Linear Collider(ILC)\cite{ILC}, and
 of 3 $TeV$ to the Compact Linear Collider(CLIC)\cite{CLIC}. In many cases, the ILC can significantly improve
 the LHC measurements. If a Higgs boson is discovered, it will be crucial to determine its
 couplings with high accuracy. The
 running of the high energy and luminosity linear collider will
 open an unique window for us to reach understanding of the fundamental theory of
 particle physics. So far, many works have been contributed to studies of the Higgs boson pair
 production at the ILC, in the SM \cite{sm} and in the new physics beyond the SM \cite{bsm}.
 In this work, we will study the production
of the pair charged and neutral scalars of the LRTH model associated
with a $Z$ boson at the future high energy $e^{+}e^{-}$ linear
 colliders.\\
\indent This paper is organized as follows. In section II, we give a
briefly review of the LRTH model, and then give the relevant
couplings which are related to our calculation. Sections III and IV
are devoted to the computation of the production cross sections of
the processes $e^{+}e^{-}\rightarrow Z\phi^{+}\phi^{-}$ and
$e^{+}e^{-}\rightarrow Z\phi^{0}\phi^{0}$. Some phenomenological
 analysis are also included in the two sections. The conclusions are
given in section V. In the appendix A and B, we present the Feynman rules and formulas relevant to our calculations.\\
\noindent{\bf II. Review of the LRTH model}\\
\indent In this section we will briefly review the essential
features of the LRTH model and focusing on particle content and the
couplings relevant to our computation.\\
\indent In LRTH model, the global symmetry is $U(4)_{1}\times
U(4)_{2}$ with a locally gauged $SU(2)_{L}\times SU(2)_{R}\times
U(1)_{B-L}$ subgroup. The twin symmetry which is required to control
the quadratic divergences of the Higgs mass is identified with the
left-right symmetry which interchanges L and R, implying the gauge couplings of $SU(2)_{L}$ and $SU(2)_{R}$ are identical. \\
\indent Two Higgs fields, $H$ and $\hat{H}$, are introduced and each
transforms as $(4,1)$ and $(1,4)$ respectively under the global
symmetry. They are written as
\begin{eqnarray}
H=\left( \begin{array}{c} H_{L}\\
H_{R} \\
\end{array}  \right)\,,~~~~~~~~~~~~~~\hat{H}=\left( \begin{array}{c} \hat{H}_{L}\\
\hat{H}_{R} \\
\end{array}  \right)\,,
\end{eqnarray}
where $H_{L,R}$ and $\hat{H}_{L,R}$ are two component objects which
are charged under the $SU(2)_{L}\times SU(2)_{R}\times U(1)_{B-L}$
as
\begin{equation}
H_{L}~and~ \hat{H}_{L}: (2, 1, 1),~~~~~~~~H_{R}~ and~ \hat{H}_{R}:
(1, 2, 1).
\end{equation}
The global $U(4)_{1}(U(4)_{2})$ symmetry is spontaneously broken
down to its subgroup $U(3)_{1}(U(3)_{2})$ with non-zero vacuum
expectation values(VEV) as $\langle H\rangle=(0,0,0,f)$ and $\langle
\hat{H}\rangle=(0,0,0,\hat{f})$. Each spontaneously symmetry
breaking results in seven Nambu-Goldstone bosons. Three of six
Goldstone bosons that are charged under $SU(2)_{R}$ are eaten by the
new gauge bosons $W_{H}^{\pm}$ and $Z_{H}$, while leaves three
physical Higgs: $\phi^{0}$ and $\phi^{\pm}$. After the SM
electroweak symmetry breaking, the three additional Goldstone bosons
are eaten by the SM gauge bosons $W^{\pm}$ and $Z$. The remaining
Higgses are the SM Higgs doublet $H_{L}$ and an extra Higgs doublet
$\hat{H}_{L}=(\hat{H}_{1}^{+},\hat{H}_{2}^{0})$ that only couples to
the gauge boson sector. A residual matter parity in the model
renders the neutral Higgs $\hat{H}_{2}^{0}$ stable, and it could be
a good dark matter candidate. These Higgs bosons can couple to each
other, and also can couple to the gauge bosons. The
forms of the couplings relevant to our calculation, are given in Appendix A and B.\\
\indent As previously said, the Higgs mechanism for both $H$ and
$\hat{f}$ makes the six gauge bosons massive whereas one gauge
boson, photon, massless. There masses are expressed as:
\begin{eqnarray}
M_{A}^{2}&=&0,\\
M_{W}^{2}&=& \frac{1}{2}g^{2}f^{2}\sin^{2}x, \\
M_{W_{H}}^{2}&=& \frac{1}{2}g^{2}(\hat{f}^{2}+f^{2}\cos^{2}x),\\
M_{Z}^{2}&=&\frac{g^{2}+2g'^{2}}{g^{2}+g'^{2}}\frac{2M_{W}^{2}M_{W_{H}}^{2}}{M_{W}^{2}+M_{W_{H}}^{2}+\sqrt{(M_{W_{H}}^{2}-M_{W}^{2})^{2}+4\frac{g'^{2}}{g^{2}+g'^{2}}M_{W_{H}}^{2}M_{W}^{2}}},\\
M_{Z_{H}}^{2}&=&
\frac{g^{2}+g'^{2}}{g^{2}}(M_{W}^{2}+M_{W_{H}}^{2})-M_{Z}^{2},\end{eqnarray}
where $x=v/(\sqrt{2}f)$ and $v$ is the electroweak scale, the values
of $f$ and $\hat{f}$ will be bounded by electroweak precision
measurements. In addition, $f$ and $\hat{f}$ are interconnected once
we set $v=246GeV$. The heavy gauge bosons $Z_{H}$ and $W_{H}^{\pm}$
typically have masses of the order of 1 TeV. The Weinberg angle in
the LRTH model are defined as:
\begin{eqnarray}
S_{W}&=& \sin\theta_{W}= \frac{g'}{\sqrt{g^{2}+2g'^{2}}},\\
C_{W}&=& \cos\theta_{W}= \sqrt{\frac{g^{2}+g'^{2}}{g^{2}+2g'^{2}}}.
\end{eqnarray}
The unit of the electric charge is then given by
\begin{eqnarray}
e=gS_{W}=\frac{gg'}{\sqrt{g^{2}+2g'^{2}}}.
\end{eqnarray}
At the leading order, the mixing angles for left-handed and
right-handed fermions are
\begin{eqnarray}
s_{L}&\simeq& \frac{M}{m_{T}}\sin x,\\
s_{R}&\simeq& \frac{M}{m_{T}}(1+\sin^{2}x),
\end{eqnarray}
where $M$ is the mass parameter essential to the mixing between the
SM-like top quark and the heavy top quark. \\
\indent It has been shown that the charged Higgs $\phi^{\pm}$
dominantly decay into $tb$ for $M>10 GeV$ \cite{Hock}. In Table I,
we list the main decay branching ratios of the charged Higgs bosons
in the LRTH model. One can see that, the branching ratio
$\phi^{+}\rightarrow t\bar{b}$ is larger than $50\%$ in wide range
of
the parameter space of the LRTH model. \\
 \null\noindent ~~{Table I:} The decay
branching ratio $\phi^{+}\rightarrow t\bar{b}$ in the LRTH model for
M=50GeV and 150 GeV. \vspace{0.1in}
\begin{center}
\doublerulesep 0.8pt \tabcolsep 0.1in
\begin{tabular}{|c|c|c|c|c|c|c|c|c|c|c|}\hline \hline f (GeV)&500&600&700&800&900&1000&1200&1500\\
\hline $M=50
GeV$&$90.1\%$&$85.8\%$&$81.3\%$&$76.7\%$&$70\%$&$67.5\%$&$58.8\%$&$52.4\%$
\\ \hline $M=150
GeV$&$98.7\%$&$98.1\%$&$97.4\%$&$96.6\%$&$95.8\%$&$94.8\%$&$92.7\%$&$89.1\%$
\\ \hline \hline \end{tabular}
\end {center}

\indent At the leading order, the total decay width $\Gamma_{Z_{H}}$
of the heavy gauge boson $Z_{H}$ is dominated by
$q\bar{q}(q=u,d,s,c)$ and $b\bar{b}$, which can be written as: $\Gamma_{Z_{H}}\simeq 0.02M_{Z_{H}}$\cite{Hock}.\\
 \noindent{\bf III. The process of $e^{+}e^{-}\rightarrow Z\phi^{+}\phi^{-}$}\\
\begin{figure}[ht]
\begin{center}
\epsfig{file=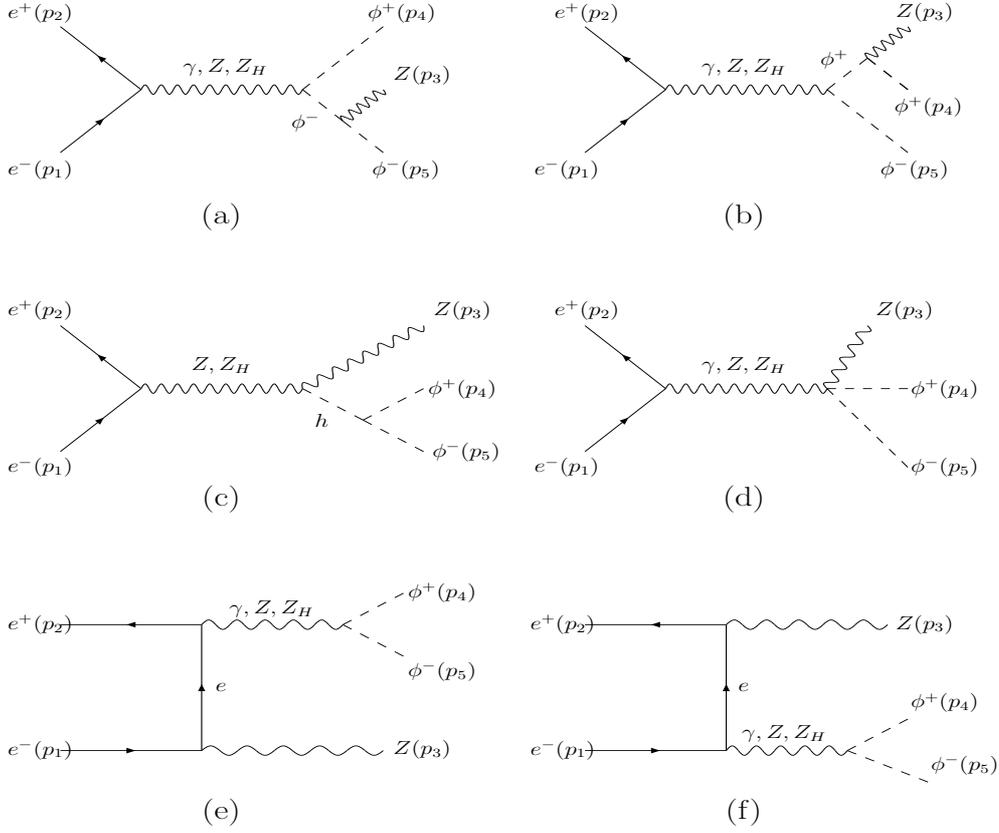,width=450pt,height=500pt} \vspace{-5.5cm}
\caption{\small Feynman diagrams of the process
$e^{+}e^{-}\rightarrow Z\phi^{+}\phi^{-}$ in the left-right twin
Higgs model.} \label{fig1}
\end{center}
\end{figure}
 \indent In LRTH model, the
charged Higgs bosons pair $\phi^{+}\phi^{-}$ can
 be produced via $e^{+}e^{-}$ annihilation associated with a $Z$
 boson as shown in figure 1. The relevant Feynman rules are given in appendix A. The invariant production amplitudes of the
 process $e^{+}e^{-}\rightarrow Z\phi^{+}\phi^{-}$ can be written
 as:
\begin{equation}
\mathcal{M}_{1}= \mathcal{M}_{a}+ \mathcal{M}_{b}+
\mathcal{M}_{c}+M_{d}+ \mathcal{M}_{e}+ \mathcal{M}_{f},
 \end{equation}
 with
 \begin{eqnarray}
  \mathcal{M}_{a}&=&\sum_{i=1}^{3}\bar{v_{e}}(p_{2})i\gamma_{\mu}(g_{V_{i}}+g_{A_{i}}\gamma_{5})u_{e}(p_{1})\frac{ig^{\mu\nu}}{p_{12}^{2}-M_{i}^{2}+iM_{i}\Gamma_{i}}(iE_{i}^{\phi^{+}\phi^{-}})(p_{4}-p_{5}-p_{3})_{\nu}\nonumber\\
 & & \times\frac{i}{p_{35}^{2}-M_{\Phi}^{2}}(iE_{Z}^{\phi^{+}\phi^{-}})(-p_{3}-2p_{5})_{\alpha}\epsilon^{\alpha}(p_{3}),\\
 \mathcal{M}_{b}&=&\sum_{i=1}^{3}\bar{v_{e}}(p_{2})i\gamma_{\mu}(g_{V_{i}}+g_{A_{i}}\gamma_{5})u_{e}(p_{1})\frac{ig^{\mu\nu}}{p_{12}^{2}-M_{i}^{2}+iM_{i}\Gamma_{i}}(iE_{i}^{\phi^{+}\phi^{-}})(p_{3}+p_{4}-p_{5})_{\nu}\nonumber\\
 & & \times\frac{i}{p_{34}^{2}-M_{\Phi}^{2}}(iE_{i}^{\phi^{+}\phi^{-}})(2p_{4}+p_{3})_{\alpha}\epsilon^{\alpha}(p_{3}),\\
 \mathcal{M}_{c}&=&\sum_{i=2}^{3}\bar{v_{e}}(p_{2})i\gamma_{\mu}(g_{V_{i}}+g_{A_{i}}\gamma_{5})u_{e}(p_{1})\frac{ig^{\mu\nu}}{p_{12}^{2}-M_{i}^{2}+iM_{i}\Gamma_{i}}(iV_{i}^{HZ_{i}Z})g_{\nu\alpha}\nonumber\\
 & & \times\frac{i}{p_{45}^{2}-M_{h}^{2}}V^{\phi^{+}\phi^{-}h}\epsilon^{\alpha}(p_{3}),\\
 \mathcal{M}_{d}&=&\sum_{i=1}^{3}\bar{v_{e}}(p_{2})i\gamma_{\mu}(g_{V_{i}}+g_{A_{i}}\gamma_{5})u_{e}(p_{1})\frac{ig^{\mu\nu}}{p_{12}^{2}-M_{i}^{2}+iM_{i}\Gamma_{i}}(iC_{i1}^{\phi^{+}\phi^{-}})g_{\nu\alpha}\epsilon^{\alpha}(p_{3}),\\
 \mathcal{M}_{e}&=&\sum_{i=1}^{3}\bar{v_{e}}(p_{2})i\gamma_{\mu}(g_{V_{i}}+g_{A_{i}}\gamma_{5})\frac{ig_{\mu\nu}}{p^{2}_{45}-M_{i}^{2}+iM_{i}\Gamma_{i}}(iE_{i}^{\phi^{+}\phi^{-}})(p_{4}-p_{5})_{\nu}\nonumber\\
 & & \times
 i\frac{\pslash_{3}-\pslash_{1}}{(p_{3}-p_{1})^{2}}\gamma_{\alpha}(g_{V_{2}}+g_{A_{2}}\gamma_{5})\epsilon^{\alpha}(p_{3})u_{e}(p_{1}),\\
 \mathcal{M}_{f}&=&\sum_{i=1}^{3}\bar{v_{e}}(p_{2})i\gamma_{\mu}(g_{V_{2}}+g_{A_{2}}\gamma_{5})\epsilon^{\mu}(p_{3})i\frac{\pslash_{3}-\pslash_{2}}{(p_{3}-p_{2})^{2}}i\gamma_{\nu}(g_{V_{i}}+g_{A_{i}}\gamma_{5})u_{e}(p_{1})\nonumber\\
 & & \times\frac{ig_{\nu
 \alpha}}{p^{2}_{45}-M_{i}^{2}+iM_{i}\Gamma_{i}}(iE_{i}^{\phi^{+}\phi^{-}})(p_{4}-p_{5})_{\alpha}.
 \end{eqnarray}
  Where $p_{12}$ is the momentum of the propagator, which is the sum of the incoming momentums $p_{1}$ and $p_{2}$.
  $M_{\Phi}$ denote the mass of $\phi^{-}$, $\epsilon_{\alpha}(p_{3})$ is the polarization vector of the $Z$ boson,
 $p_{4}$ and $p_{5}$ denote the momenta of outgoing charged Higgs bosons $\phi^{+}$ and
 $\phi^{-}$. $\Gamma_{i}$
represents the gauge bosons total decay width. \\
 \indent With the above production amplitudes, we can obtain the
production cross section directly. In the calculation of the cross
section, instead of calculating the square of the amplitudes
analytically, we calculate the amplitudes numerically by using the
method of the references \cite{HZ} which can greatly simplify our
calculation. Finally we also use the CalcHEP \cite{calchep} packages to check our results.\\
\indent In performing the numerical calculations, we take the
 SM input parameters as $\alpha_{e}$=1/128.8, $m_{Z}=91.1876GeV$,  $m_{h}=120 GeV$, $s_{W}^{2}$=0.2226 and
 $\Gamma_{Z}=2.495GeV$\cite{data}. The free LRTH model parameters are $f$, the heavy gauge boson mass $M_{Z_{H}}$,
 and the mass of the charged Higgs boson $M_{\phi}$. Taking into account the
precision electroweak constraints on the parameter space, the
symmetry breaking scales $f$ is allowed in the range of $500GeV \sim
1500 GeV$. It has been shown $M_{\phi^{-}}$ is allowed to be in the
range of a few hundred GeV depending on the model \cite{Hock}. As
numerical estimation, we will assume that the charged Higgs bosons
mass $M_{\phi}$ and the $Z_{H}$ mass
$M_{Z_{H}}$ are in the ranges of $150 GeV \sim 400 GeV$ and $1 TeV \sim 3TeV$, respectively.\\
\begin{figure}[t]
\begin{center}
\scalebox{0.85}{\epsfig{file=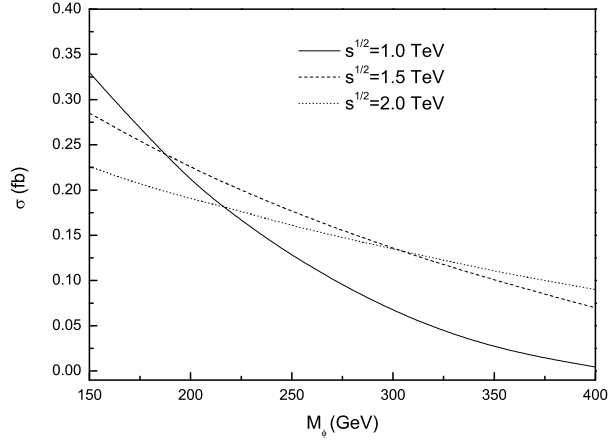}}
\end{center}
\caption{\small The production cross section $\sigma$ versus charged
Higgs mass $M_{\phi}$ for $f=1000 GeV$, $M_{Z_{H}}=3.0 TeV$ and
three values of $\sqrt{s}$.}
\end{figure}
\indent In Fig.2, we plot the cross section $\sigma$ the process
$e^{+}e^{-}\rightarrow Z\phi^{+}\phi^{-}$ as a function of the mass
parameter $M_{\phi}$ for $f=1000 GeV$, $M_{Z_{H}}=3.0 TeV$ and three
values of the center of mass energy. The plots show that the cross
section $\sigma$ decreases with $M_{\phi}$ increasing, due to phase
space suppression. The change of the cross section with $\sqrt{s}$
is not monotonic because the influence of $\sqrt{s}$ on the phase
space and the gauge boson propagators is inverse. In this case, the
production rate is at the level of $10^{-1}fb$. For $\sqrt{s}=1.0
TeV$ and $150 GeV\leq M_{\phi}\leq 300GeV$, the value of $\sigma$ is
in the range of $0.06fb\sim 0.32fb$. If we assume that the future
ILC experiment with $\sqrt{s}$=1.0 TeV has a yearly integrated
luminosity of
$500fb^{-1}$, then there will be $10^{2}-10^{3}$ signal events generated at the ILC. \\
\begin{figure}[t]
\begin{center}
\scalebox{0.85}{\epsfig{file=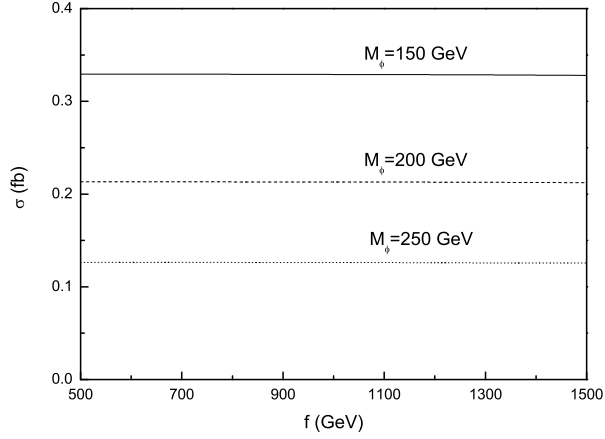}}
\end{center}
\caption{\small The production cross section $\sigma$ as a function
of the parameter $f$ for $\sqrt{s}=1.0 TeV$, $M_{Z_{H}}=3.0 TeV$ and
various values of $M_{\phi}$.}
\end{figure}
\indent To see the influence of the scalar parameter $f$ on the
cross section, in Fig. 3 we plot the cross section $\sigma$ as a
function of $f$ for $\sqrt{s}=1.0 TeV$ and three values of
$M_{\phi}=150, 200$ and $300GeV$, respectively. From Fig. 3, one can
see that the cross section is not sensitive to $f$. This is because
the contributions come from Fig .1(c) to the production cross
section of the process $e^{+}e^{-}\rightarrow Z\phi^{+}\phi^{-}$ is
suppressed by a factor of $(x^{2}/2f^{4})$, which is included in the
scalar self-interactions $\phi^{+}\phi^{-}h$. So, in our
calculation, we can safely neglect the effect of different values of
$f$ to the cross section.\\
\begin{figure}[t]
\begin{center}
\scalebox{0.75}{\epsfig{file=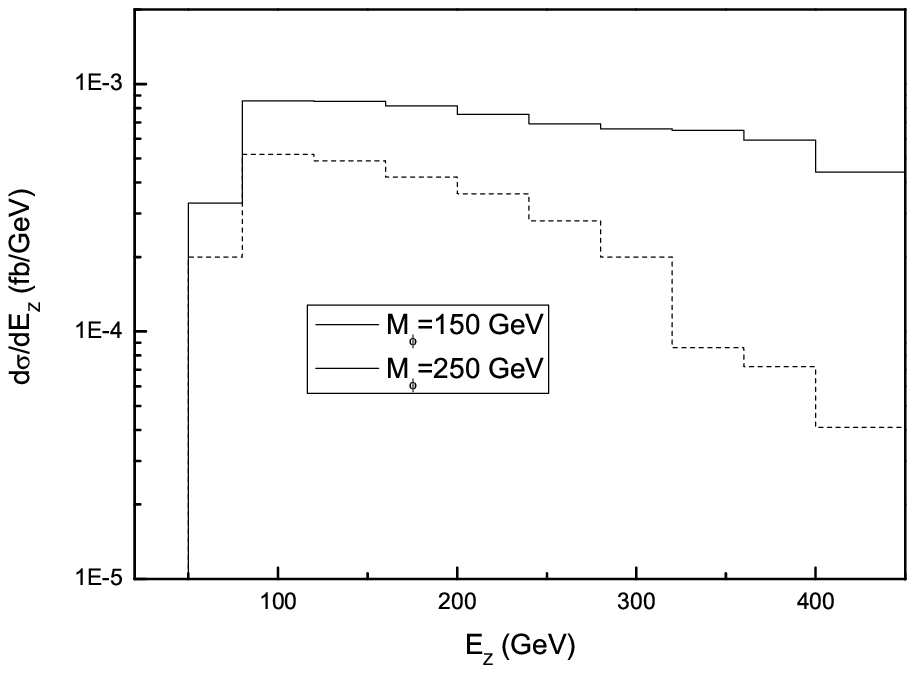}\epsfig{file=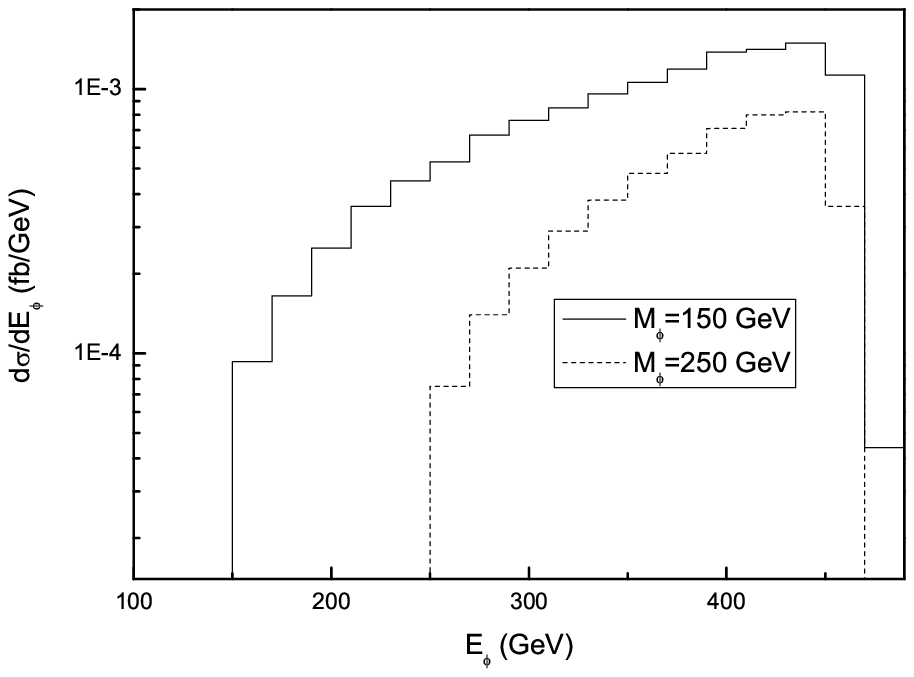}}
\end{center}
\caption{\small Differential cross sections versus $E_{Z}$ and
$E_{\phi^{-}}$ graphs for $\sqrt{s}=1.0 TeV$, $f=1000 GeV$,
$M_{Z_{H}}=2.0 TeV$ and two values of $M_{\phi}$. (a) for $Z$ boson,
(b) for charged Higgs boson $\phi^{-}$.}
\end{figure}
\begin{figure}[b]
\begin{center}
\scalebox{0.75}{\epsfig{file=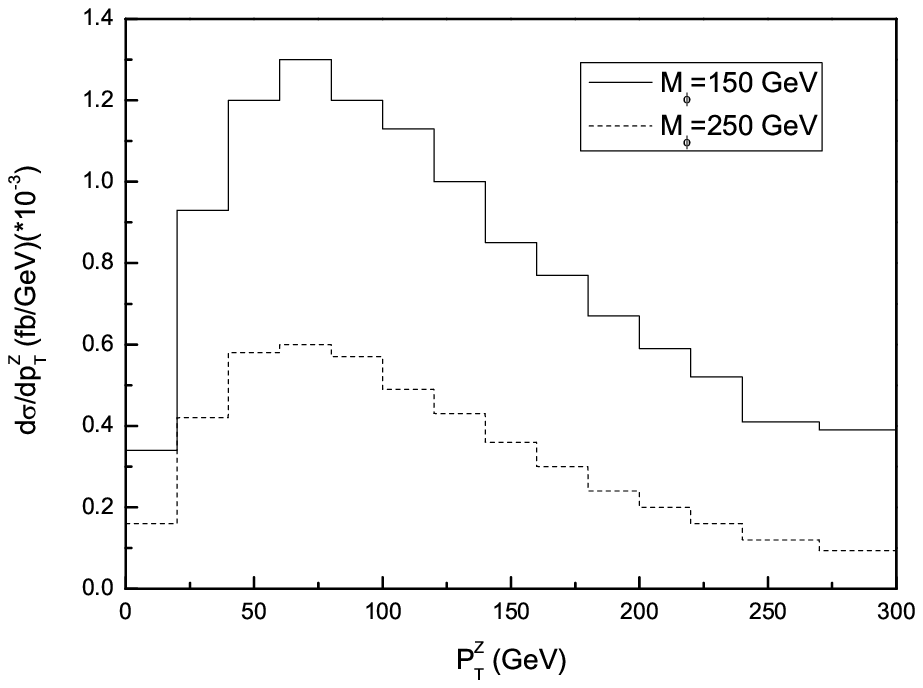}\epsfig{file=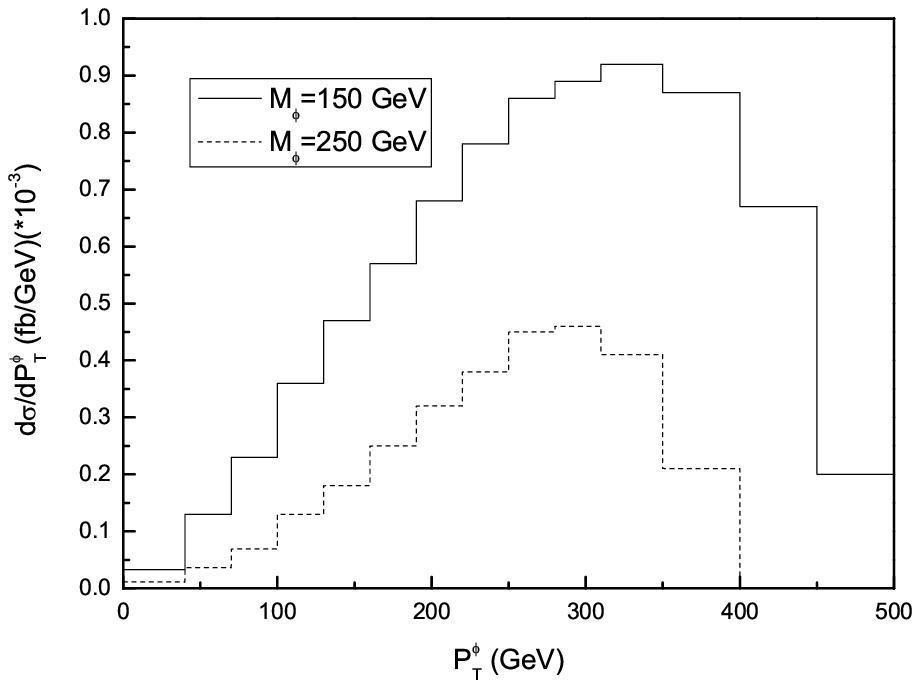}}
\end{center}
\caption{\small Distributions of the transverse momenta of $Z$ boson
and charged Higgs boson $\phi^{-}$ for the $e^{+}e^{-}\rightarrow
Z\phi^{+}\phi^{-}$ process with $\sqrt{s}=1.0 TeV$ and two values of
$M_{\phi}$. (a) for $Z$ boson, (b) for charged Higgs boson
$\phi^{-}$. }
\end{figure}
\indent The distributions of the energy of $Z$ boson and charged
Higgs boson $\phi^{-}$ are shown in Fig. 4 for $\sqrt{s}=1.0 TeV$
and $M_{\phi}=150 GeV$ and $250 GeV$, respectively. We can see from
the Fig.4(a) that the peak values of differential cross sections are
obtained at the order of $10^{-3}\frac{fb}{GeV}$ for low $E_{Z}$
values, $E_{Z}\sim 100 GeV$. Meanwhile, the values of $E_{\phi^{-}}$
ranging from $400 GeV$ to $450 GeV$ make the main contribution to
the cross section of $e^{+}e^{-}\rightarrow Z\phi^{+}\phi^{-}$.\\
\indent In Fig.5, we provide the distributions of transverse momenta
of $p_{T}^{Z}$ and $p_{T}^{\phi^{-}}$ with $\sqrt{s}=1.0 TeV$ and
two values of the charged Higgs bosons mass. From these two figures
we can see that, the significant regions of $p_{T}$ for $Z$ boson
and charged Higgs boson $\phi^{-}$ are in the regions of $50 GeV\sim
100 GeV$, and $250 GeV\sim 350 GeV$, respectively. \\

\begin{figure}[ht]
\begin{center}
\scalebox{0.75}{\epsfig{file=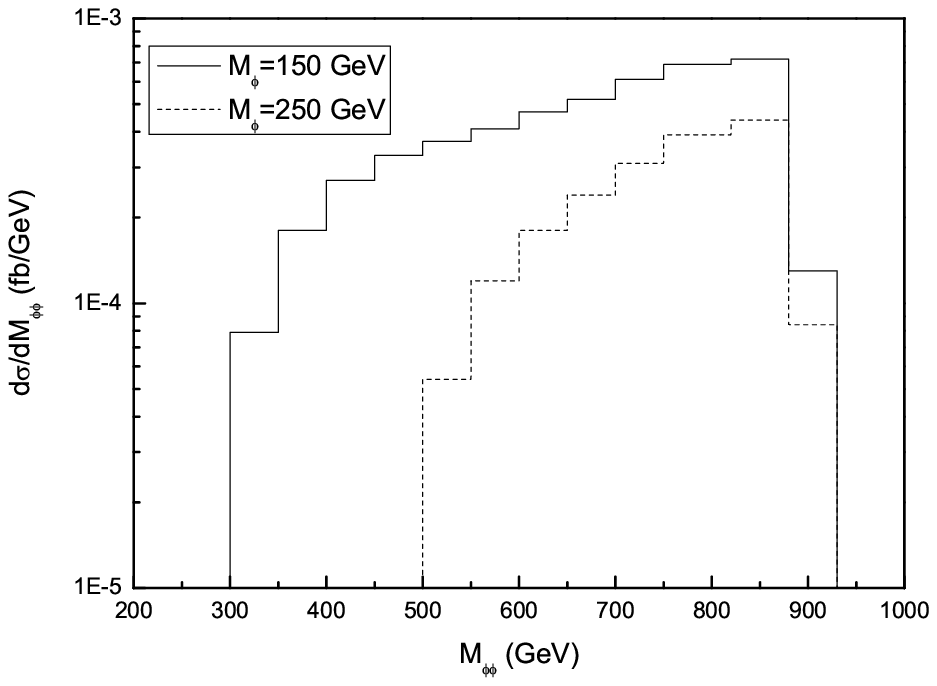}\epsfig{file=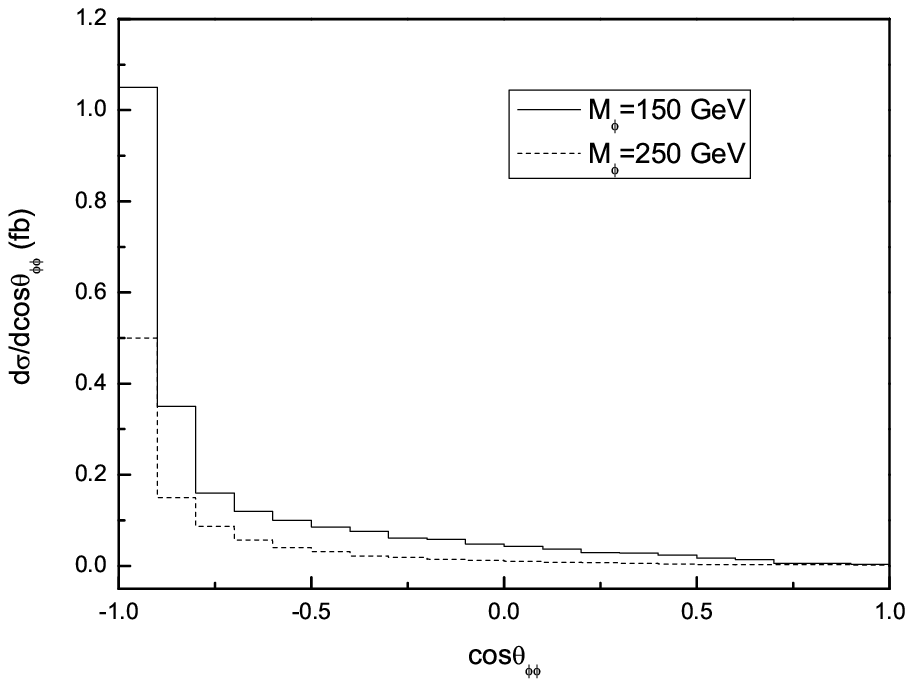}}
\end{center}
\caption{\small (a) Distributions of the invariant mass of charged
Higgs bosons pair with $\sqrt{s}=1.0 TeV$ and two values of
$M_{\phi}$. (b) Differential cross sections of the cosine of the
angle between the produced charged Higgs bosons pair with
$\sqrt{s}=1.0 TeV$ and two values of $M_{\phi}$. }
\end{figure}
\begin{figure}[ht]
\begin{center}
\scalebox{0.75}{\epsfig{file=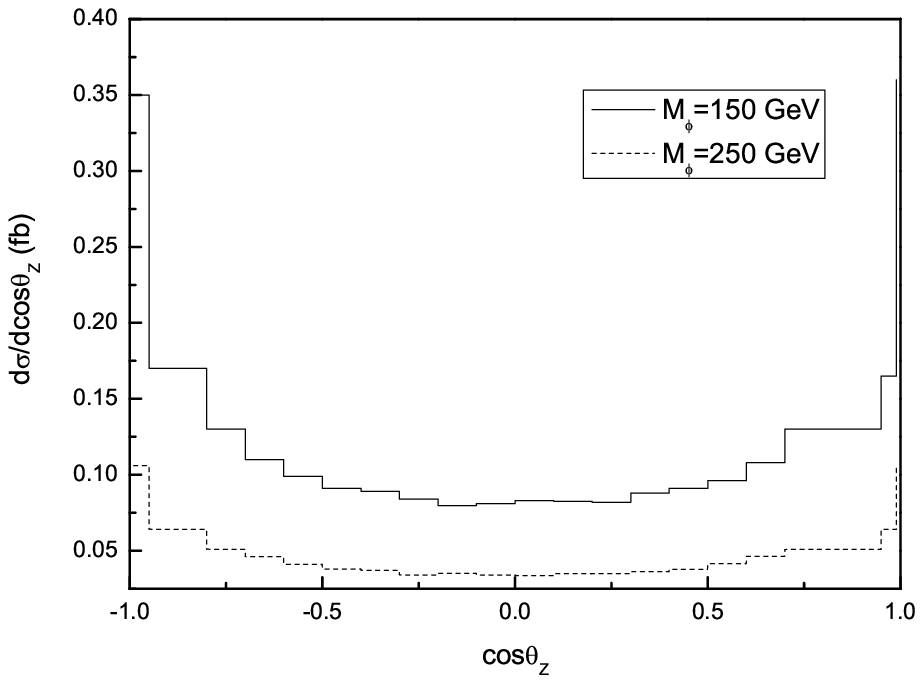}\epsfig{file=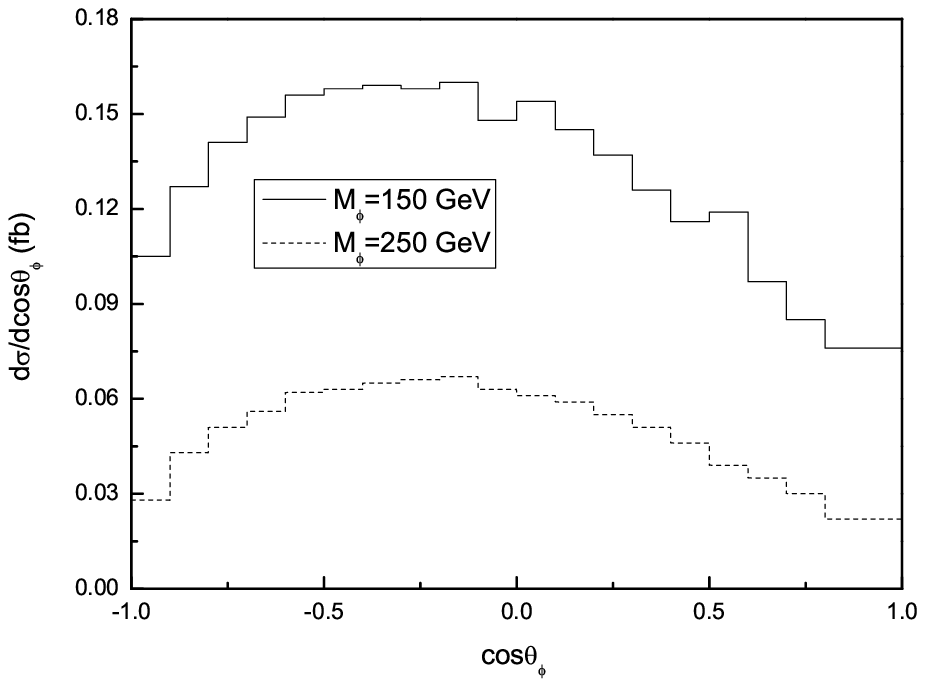}}
\end{center}
\caption{\small Distributions of the cosine of the $Z$ boson (
charged Higgs boson $\phi^{-}$) production angle with respect to
z-axis for the $e^{+}e^{-}\rightarrow Z\phi^{+}\phi^{-}$ process
with $\sqrt{s}=1.0 TeV$ and two values of $M_{\phi}$. (a) for
$\frac{d\sigma}{d\cos \theta_{Z}}$, (b) for $\frac{d\sigma}{d\cos
\theta_{\phi^{-}}}$. }
\end{figure}
\indent The distribution of the charged Higgs bosons pair invariant
mass $M_{\phi\phi}$ is shown in Fig. 6(a), and the differential
cross section of the cosine of the angle between the produced
charged Higgs bosons pair is shown in Fig.6(b) where we take two
values of $M_{\phi}$ and $\sqrt{s}=1.0 TeV$. We can see from the
Fig.6(a) that the relatively large $M_{\phi\phi}$ region (from $700
GeV$ to $880 GeV$) make the main contribution to the production
cross section of $e^{+}e^{-}\rightarrow Z\phi^{+}\phi^{-}$. Fig.6(b)
shows the distribution of cosine of the angle between the produced
charged Higgs bosons pair. We can see from the figure that the
produced charged Higgs boson pair prefer to go out almost back to
back, that leads to the $M_{\phi\phi}$ having the tendency to
distribution in
large value region.\\
\indent We take the orientation of incoming electron as the z-axis.
The $\theta_{Z}$ (or $\theta_{\phi^{-}}$) is defined as the
$Z$-boson (or charged Higgs boson $\phi^{-}$) production angle with
respect to the z-axis. In Fig.7(a,b) we present the distributions of
cosines of the pole angles of $Z$-boson and charged Higgs boson
$\phi^{-}$ ($\cos \theta_{Z}$ and $\cos \theta_{\phi^{-}}$)
respectively, in conditions of $\sqrt{s}=1.0 TeV$ and two values of
$M_{\phi}$. From Fig.7(a), it can be seen that the outgoing
$Z$-boson is symmetry in the forward and background hemisphere
region, while Fig.7(b) demonstrates that the significant regions of
$\cos \theta_{\phi^{-}}$ for charged Higgs boson $\phi^{-}$ are
rather large.\\
\begin{figure}[b]
\begin{center}
\scalebox{0.75}{\epsfig{file=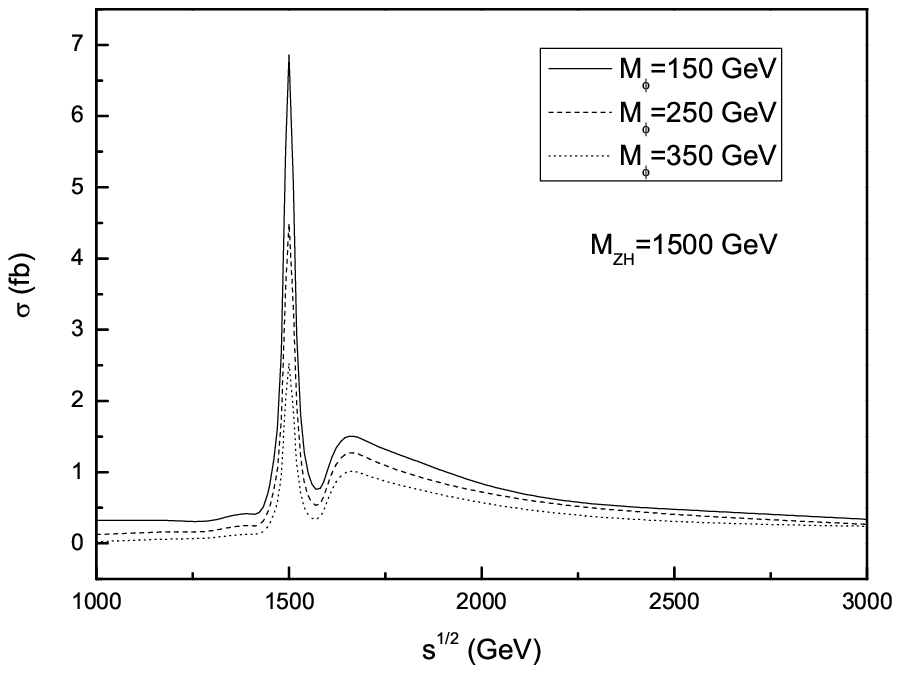}\epsfig{file=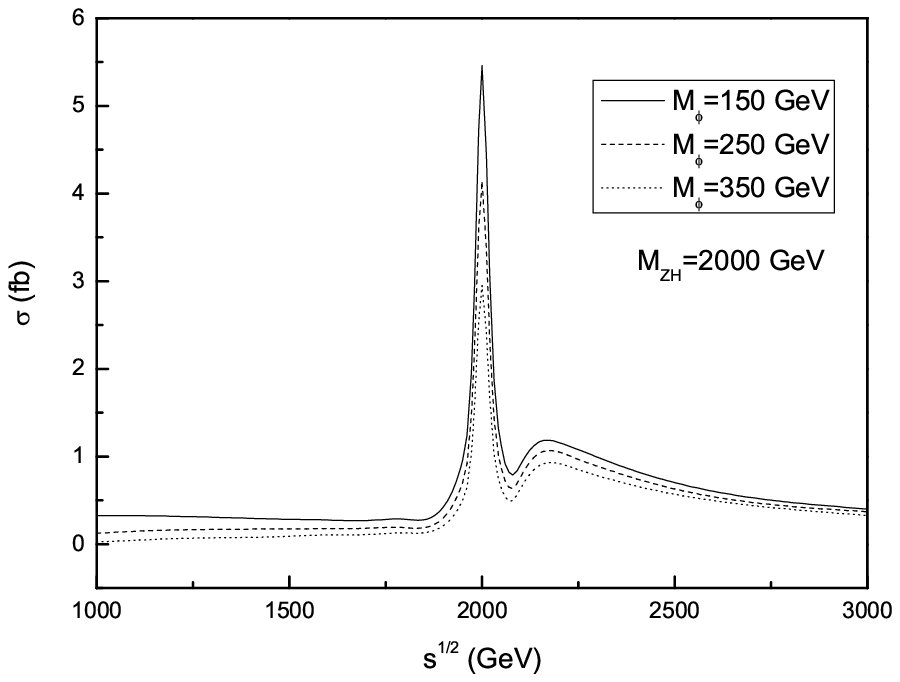}}
\end{center}
\caption{\small The production cross section $\sigma$ versus
$\sqrt{s}$ for $M_{Z_{H}}=1500 GeV$ (left), and $M_{Z_{H}}=2000
GeV$(right), and three different values of the charged Higgs mass.}
\end{figure}
  \indent To show the influence of c.m. energy on the cross section
$\sigma$, in Fig.8, we give the cross section plots as the function
of $\sqrt{s}$ with fixed $M_{Z_{H}}$ and three values of the charged
Higgs bosons mass $M_{\Phi}$. From Fig.8, we can see that the cross
section $\sigma$ resonance emerges when the $Z_{H}$ mass $M_{Z_{H}}$
approaches the c.m. energy $\sqrt{s}$. The resonance values of the
$\sigma$ decrease as $M_{Z_{H}}$ increase. For $M_{Z_{H}}=1500 GeV$
and $M_{Z_{H}}=1500 GeV$ and $2000GeV$, the cross section $\sigma$
can reach $6.9 fb$ and $5.5 fb$, respectively. Therefor, if we
assume the integrated luminosity for the ILC is $500 fb^{-1}$, there
will be thousands of $Z\phi^{+}\phi^{-}$ signal events generated at
the ILC. \\
\begin{figure}[b]
\begin{center}
\scalebox{0.75}{\epsfig{file=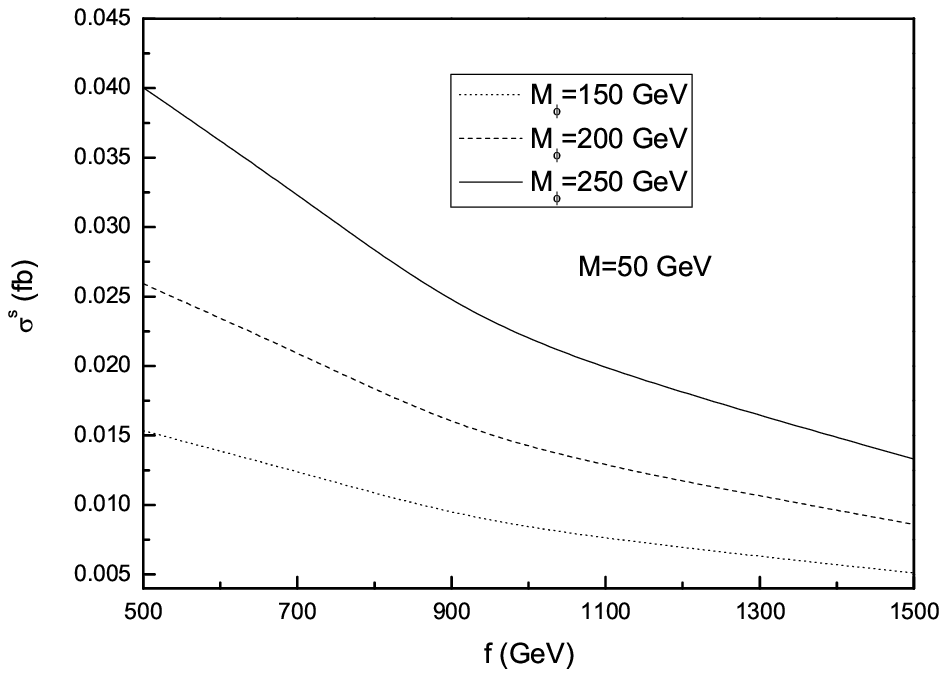}\epsfig{file=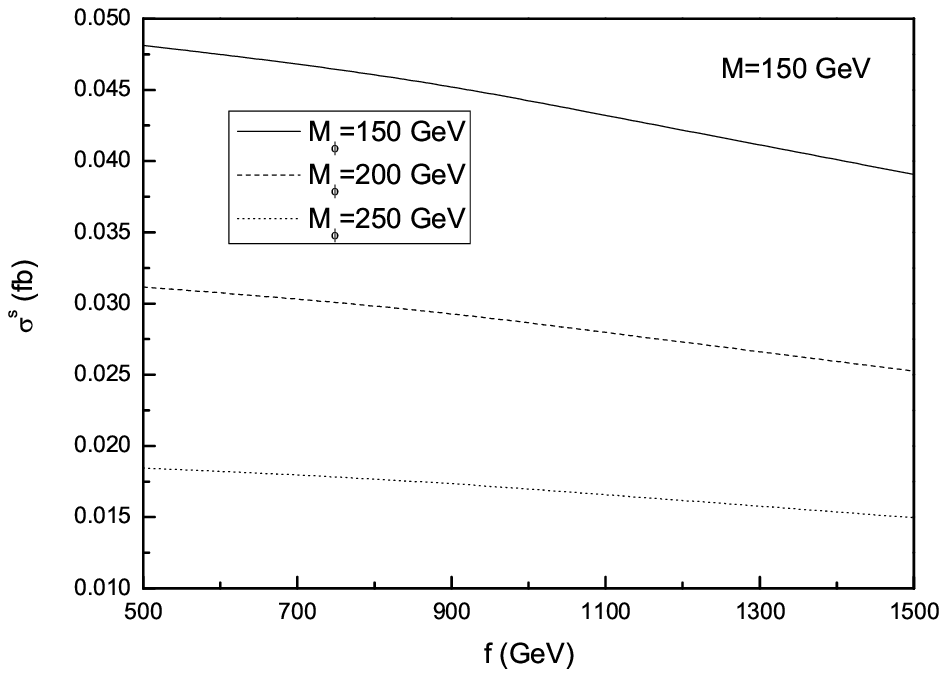}}
\end{center}
\caption{\small The production rate of the
$t\bar{t}b\bar{b}b\bar{b}$ final state as a function of the
parameter $f$ for $\sqrt{s}=1.0 TeV$, $M=50 GeV$ (left), and $M=150
GeV$(right), and various values of $M_{\phi}$.}
\end{figure}
\indent Considering the subsequent decay of $\phi^{+}\rightarrow
t\bar{b}$, $t\rightarrow W^{+}b\rightarrow l^{+}\nu b$, the
characteristic signal final state of $Z\phi^{+}\phi^{-}$, including
four $b$ jets + four charged lepton ($e$ or $\mu$) +missing $E_{T}$
and six jet $q\bar{q}b\bar{b}b\bar{b}$ + two lepton +missing
$E_{T}$, which are coming from the $Z$ boson decaying to a charged
leptons and $q\bar{q}$, respectively. The $Z$ boson in the final
state gives an unambiguous event identification via its leptonic
decay. In the case of $Z\rightarrow b\bar{b}$, the production rate
of the $t\bar{t}b\bar{b}b\bar{b}$ final state can be easily
estimated $\sigma^{s}\approx \sigma\times [Br(Z\rightarrow
b\bar{b})\times Br(\phi^{+}\rightarrow t\bar{b})\times
Br(\phi^{-}\rightarrow \bar{t}b)]$. The numerical results are shown
in Fig.9. One can see from this figure that, with reasonable values
of the free parameters of the LRTH model, the production rate can
reach $0.04 fb$. However, its value decreases quickly as the mass of
charged Higgs bosons $M_{\phi}$ increases. The main backgrounds for
the $t\bar{t}b\bar{b}b\bar{b}$ final state come from the SM
processes $e^{+}e^{-}\rightarrow t\bar{t}hh$, $e^{+}e^{-}\rightarrow
t\bar{t}Zh$ and $e^{+}e^{-}\rightarrow t\bar{t}ZZ$ with
$Z\rightarrow b\bar{b}$ and $h\rightarrow b\bar{b}$, continuum
$t\bar{t}b\bar{b}b\bar{b}$ production. For $\sqrt{s}=1.0 TeV$, the
total production rate of the $t\bar{t}b\bar{b}b\bar{b}$ backgrounds
is estimated to be about $0.01 fb$. Thus, it may be possible to
extract the signals from the backgrounds in the reasonable parameter
spaces in the LRTH model. In addition, the reconstruction of $W$,
$t$, and $\phi^{\pm}$ can be used to discriminate the signal from
the background. Certainly, detailed confirmation of the
observability of the signals generated by the process
$e^{+}e^{-}\rightarrow Z\phi^{+}\phi^{-}$ would require Monte-Carlo
simulations of the signals and backgrounds, which is beyond the
scope of this paper.
 \\
\noindent{\bf IV. The process of $e^{+}e^{-}\rightarrow Z\phi^{0}\phi^{0}$}\\
\indent With the couplings $\phi^{0}\phi^{0}h$, $h\phi^{0}Z_{i}$,
and $\phi^{0}\phi^{0}ZZ_{i}$, the processes $e^{+}e^{-}\rightarrow
Z\phi^{0}\phi^{0}$ can be induced at tree-level. The Feynman
diagrams of these process are shown in Fig.10. The amplitudes of the
process $e^{+}e^{-}\rightarrow Z\phi^{0}\phi^{0}$ can be written as:
\begin{figure}[ht]
\begin{center}
\epsfig{file=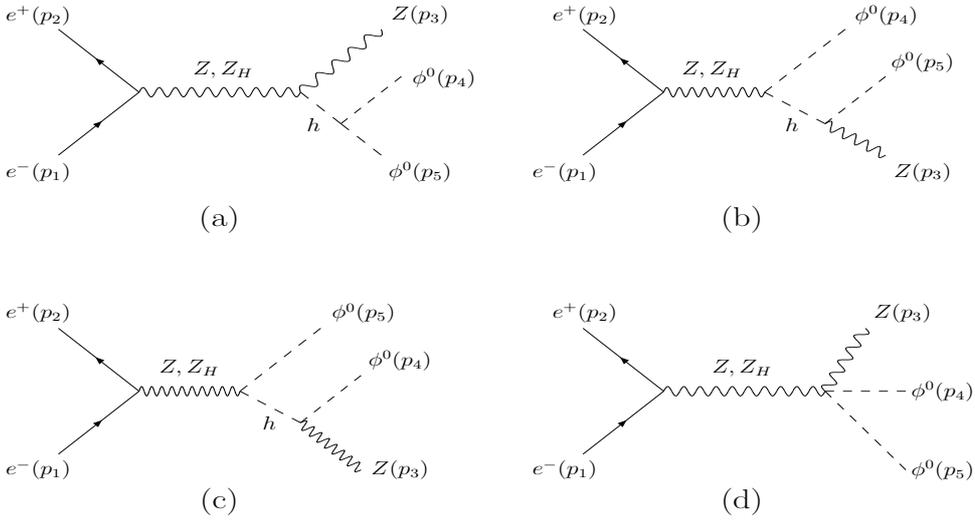,width=450pt,height=500pt} \vspace{-9cm}
\caption{\small Feynman diagrams of the process
$e^{+}e^{-}\rightarrow Z\phi^{0}\phi^{0}$ in the left-right twin
Higgs model.} \label{fig1}
\end{center}
\end{figure}
\begin{equation}
  \mathcal{M}_{2}= \mathcal{M}_{a}+ \mathcal{M}_{b}+ \mathcal{M}_{c}+ \mathcal{M}_{d},
 \end{equation}
 with
 \begin{eqnarray}
  \mathcal{M}_{a}&=&\sum_{i=2}^{3}\bar{v_{e}}(p_{2})i\gamma_{\mu}(g_{V_{i}}+g_{A_{i}}\gamma_{5})u_{e}(p_{1})\frac{ig^{\mu\nu}}{p_{12}^{2}-M_{i}^{2}+iM_{i}\Gamma_{i}}(iV^{hZ_{i}Z})g_{\nu\alpha}\nonumber\\
 & & \times\epsilon^{\alpha}(p_{3})\frac{i}{(p_{4}+p_{5})^{2}-M_{h}^{2}}V_{h\phi^{0} \phi^{0}},\\
 \mathcal{M}_{b}&=&\sum_{i=2}^{3}\bar{v_{e}}(p_{2})i\gamma_{\mu}(g_{V_{i}}+g_{A_{i}}\gamma_{5})u_{e}(p_{1})\frac{ig^{\mu\nu}}{p_{12}^{2}-M_{i}^{2}+iM_{i}\Gamma_{i}}(iV_{h\phi^{0}Z_{i}})p_{12_{\nu}}\nonumber\\
 & & \times\frac{i}{(p_{3}+p_{5})^{2}-M_{h}^{2}}(iV_{h\phi^{0}Z})p_{3_{\alpha}}\epsilon^{\alpha}(p_{3}),\\
 \mathcal{M}_{c}&=&\sum_{i=2}^{3}\bar{v_{e}}(p_{2})i\gamma_{\mu}(g_{V_{i}}+g_{A_{i}}\gamma_{5})u_{e}(p_{1})\frac{ig^{\mu\nu}}{p_{12}^{2}-M_{i}^{2}+iM_{i}\Gamma_{i}}(iV_{h\phi^{0}Z_{i}})p_{12_{\nu}}\nonumber\\
 & & \times\frac{i}{(p_{3}+p_{4})^{2}-M_{h}^{2}}(iV_{h\phi^{0}Z})p_{3_{\alpha}}\epsilon^{\alpha}(p_{3}),\\
  \mathcal{M}_{d}&=&\sum_{i=2}^{3}\bar{v_{e}}(p_{2})i\gamma_{\mu}(g_{V_{i}}+g_{A_{i}}\gamma_{5})u_{e}(p_{1})\frac{ig^{\mu\nu}}{p_{12}^{2}-M_{i}^{2}+iM_{i}\Gamma_{i}}(iC_{i1}^{\phi^{0}\phi^{0}})g_{\nu\alpha}\epsilon^{\alpha}(p_{3}).
 \end{eqnarray}
 \begin{figure}[b]
\begin{center}
\scalebox{0.85}{\epsfig{file=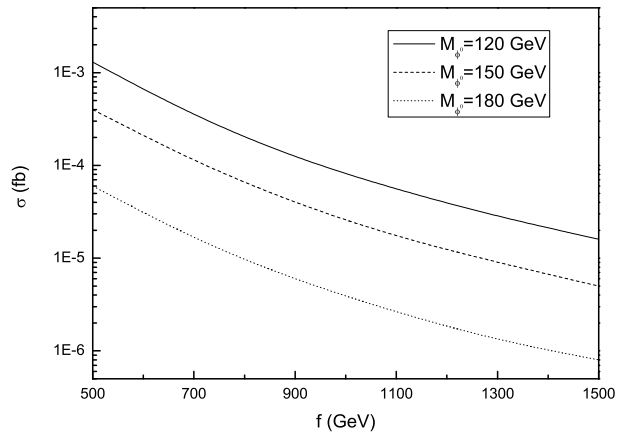}}
\end{center}
\caption{\small The production cross section $\sigma$ of
$e^{+}e^{-}\rightarrow Z\phi^{0}\phi^{0}$ versus $f$ for
$\sqrt{s}=0.5 TeV$, $M_{Z_{H}}=3.0 TeV$ and various values of
$M_{\phi^{0}}$.}
\end{figure}
 \indent In the framework of the LRTH model, the mass of the neutral
 Higgs boson $\phi^{0}$ can be anything below $f$ here we consider
 another possibility, in which the mass is about $150
 GeV$\cite{Hock,loop}. In our numerical estimation, we will assume that the
 neutral Higgs boson mass $M_{\phi^{0}}$ is in the range of $100 GeV- 180 GeV$. \\
 \indent From the relevant coupling constants in appendix B, we can see that
the production cross section of the process $e^{+}e^{-}\rightarrow
Z\phi^{0}\phi^{0}$ is very sensitive to the parameter $f$, which is
suppressed by the factor of $(v^{2}/2f^{2})$. In this case, we will
take the parameter $f$ and the neutral Higgs boson mass
$M_{\phi^{0}}$ as the free parameters. The numerical results of the
cross section versus the scalar parameter $f$ are shown in Fig.11.
We can see that $\sigma$ is sensitive to the parameter $f$ and the
mass of the neutral Higgs boson $M_{\phi^{0}}$. For $\sqrt{s}=0.5
TeV$, the value of the cross section $\sigma$ are smaller than
$1.3\times 10^{-3} fb$ in most of all parameter space preferred by
the electroweak precision data, which is really tiny and very
difficult
to detect in practice. \\
\begin{figure}[ht]
\begin{center}
\scalebox{0.75}{\epsfig{file=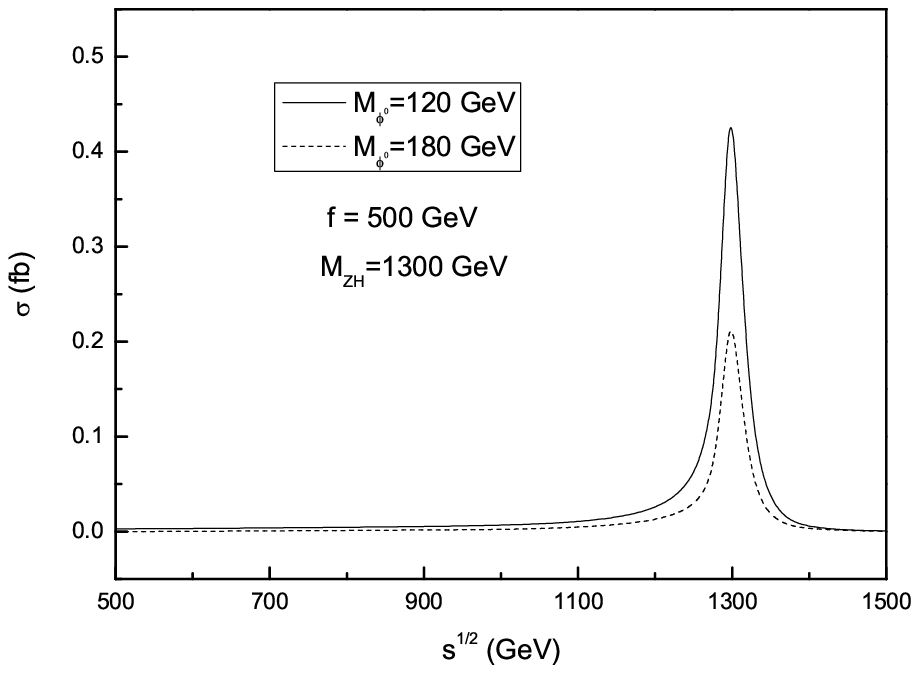}\epsfig{file=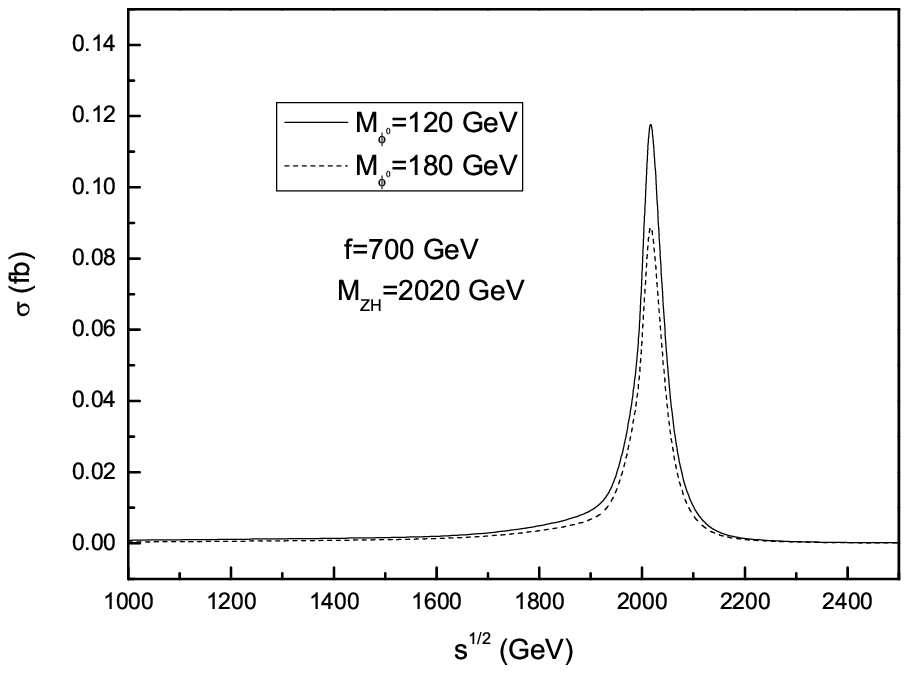}}
\end{center}
\caption{\small The production cross section $\sigma$ of
$e^{+}e^{-}\rightarrow Z\phi^{0}\phi^{0}$  versus $\sqrt{s}$ for
$f=500 GeV$(left), $f=800 GeV$right), and two typical values of
$M_{Z_{H}}$.}
\end{figure}
\indent To see the effects of the c.m. energy $\sqrt{s}$ on the
cross section $\sigma$, we plot the $\sigma$ versus $\sqrt{s}$ in
Fig.12 for two typical values of the scalar parameter $f$ and
$M_{Z_{H}}$. From Fig.12, we can see that the cross section $\sigma$
resonance emerges when the $Z_{H}$ mass $M_{Z_{H}}$ approaches the
c.m. energy $\sqrt{s}$. The resonance values of the $\sigma$ are
strongly dependent on the $M_{\phi^{0}}$ and the scalar parameter
$f$. For $M_{\phi^{0}}=120 GeV$ and $f=500 GeV$ and $700GeV$, the
cross section $\sigma$ can reach $0.5fb$ and $0.14fb$, respectively.
Therefor, if we assume the integrated luminosity for the CLIC is
$1000 fb^{-1}$, there will be tens of up to several
hundreds $Z\phi^{0}\phi^{0}$ events to be generated at the CLIC.\\
\indent Preliminary study in Ref.\cite{yue} shows that, for
$M_{\phi^{0}}=120 GeV$ and $f\leq 700 GeV$, the branching ratios
$\phi^{0}\rightarrow b\bar{b}$ are larger than $40\%$. The SM Higgs
boson $h$ has similar decay features with those of $\phi^{0}$.
Therefore, the signatures of $Z\phi^{0}\phi^{0}$ is similar to those
of $b\bar{b}hh$, $Zhh$, $ZZZ$, and $ZZh$ at the high energy
colliders. For $\sqrt{s}=2.0 TeV$, the production cross section of
the processes $e^{+}e^{-}\rightarrow b\bar{b}hh$,
$e^{+}e^{-}\rightarrow Zhh$, $e^{+}e^{-}\rightarrow ZZh$ and
$e^{+}e^{-}\rightarrow ZZZ$ are estimated to be about $0.007 fb$,
$0.054 fb$, $0.14 fb$, and $0.45 fb$, respectively. The mainly
background about six $b$ jets final state has been extensively
studied in Ref.\cite{adj}. The production rate of this kind of
signal is too small to be separated from the large
background.  \\
\noindent{\bf V. Conclusions}\\
\indent Many models of new physics beyond the SM predict the
existence of neutral or charged scalar particles. These new
particles might produce observable signatures in the current of
future high energy experiments different form the case of the SM
Higgs boson. Any visible signal from the new scalar particles will
be evidence of new physics beyond the SM. Thus, studying the new
scalar particles
production is very interesting at the ILC.\\
\indent The Left-right twin Higgs model is a concrete realization of
the twin Higgs mechanism, which predicts the existence of three
additional Higgs bosons: one neutral Higgs $\phi^{0}$ and a pair of
charged Higgs bosons $\phi^{\pm}$. In this paper, we studied the
production of a pair of charged and neutral Higgs bosons associated
with standard model gauge boson $Z$ at the ILC.  From our numerical
results, we can obtain the following conclusion: (i) For the process
$e^{+}e^{-}\rightarrow Z\phi^{+}\phi^{-}$, for $\sqrt{s}=1000 GeV$
and $150 GeV\leq M_{\Phi}\leq 400GeV$, the total production cross
section is in the range of $0.4\times 10^{-2}fb\sim 0.33fb$. If we
assume that the future ILC experiment with $\sqrt{s}$=1.0 TeV has a
yearly integrated luminosity of $500fb^{-1}$, then there will be
$10^{2}-10^{3}$ signal events generated at the ILC. Furthermore, the
$s$-channel resonance effect induced by the $Z_{H}$ gauge boson can
significantly enhance the production rate and produce enough
signals. The characteristic signal final state of 4$b$ jets + four
charged lepton +missing $E_{T}$ might be easily separated from the
SM background with a great significance. (ii) For the process
$e^{+}e^{-}\rightarrow Z\phi^{0}\phi^{0}$, the value of the cross
section $\sigma$ are smaller than $6\times 10^{-3} fb$ in most of
all parameter space preferred by the electroweak precision data.
However, for $\sqrt{s}\approx M_{Z_{H}}$, the cross section $\sigma$
can be significantly enhanced. Thus, we expect that the future ILC
experiments can be seen as an ideal tool to detect the possible
signatures of the charged and neutral Higgs bosons predicted by the
LRTH model. Even if we can not observe the signals in future ILC
experiments, at least, we can obtain the bounds on the free
parameters of the LRTH model.\\

\noindent{\bf Acknowledgments}\\
\indent We thank Shufang Su for providing the CalcHep Model Code.
This work is supported in part by the National Natural Science
Foundation of China(Grant No.10775039), and by the Foundation of
He¡¯nan Educational Committee (Grant No.2009B140003).

\newpage
\vspace{.5cm} \noindent{\bf Appendix A: The relevant coupling
constants in the process $e^{+}e^{-}\rightarrow Z\phi^{+}\phi^{-}$ }
 \vspace{0.1in}
\begin{center}
\doublerulesep 0.8pt \tabcolsep 0.1in
\begin{tabular}{||c|c|c|c||}\hline
\hline $i$ &vertices&$g_{V_{i}}$&$g_{A_{i}}$\\
\hline $1$ &$e\bar{e}\gamma$&e&0
\\ \hline $2$
&$e\bar{e}Z$&$-\frac{e}{2S_{W}C_{W}}[(-\frac{1}{2}+2S_{W}^{2})+\frac{v^{2}}{4(f^{2}+\hat{f}^{2})}\frac{S_{W}^{2}(2C_{W}^{2}-3)}{C_{W}^{4}}]$&$-\frac{e}{2S_{W}C_{W}}[\frac{1}{2}++\frac{v^{2}}{4(f^{2}+\hat{f}^{2})}\frac{S_{W}^{2}(2C_{W}^{2}-1)}{C_{W}^{4}}]$
\\ \hline $3$
&$e\bar{e}Z_{H}$&$\frac{e}{4S_{W}C_{W}\sqrt{1-2S_{W}^{2}}}(-1+4S_{W}^{2})$&$\frac{e}{4S_{W}C_{W}\sqrt{1-2S_{W}^{2}}}(2S_{W}^{2}-1)$
\\
\hline\hline
\end{tabular}
\end {center}
 \null\noindent ~~{\bf Table 1:} The vector and axial vector couplings of $e\bar{e}$ with vector bosons. Feynman rules for
 $e\bar{e}V_{i}$ vertices are given as $i\gamma_{\mu}(g_{V_{i}}+g_{A_{i}}\gamma_{5})$ \cite{Hock}. \vspace{0.1in}

\begin{center}
\doublerulesep 0.8pt \tabcolsep 0.1in
\begin{tabular}{||c|c|c||}\hline
\hline $i/j$ &vertices&$iC_{ij}^{\phi^{+}\phi^{-}}g_{\mu\nu}$\\
\hline $1/1$
&$\phi^{+}\phi^{-}A_{\mu}Z_{\nu}$&$-i\frac{2e^{2}S_{W}}{C_{W}}g_{\mu\nu}$\\
\hline $2/1$
&$\phi^{+}\phi^{-}Z_{\mu}Z_{\nu}$&$i\frac{2e^{2}S_{W}^{2}}{C_{W}^{2}}g_{\mu\nu}$
\\ \hline $3/1$
&$\phi^{+}\phi^{-}Z_{H_{\mu}}Z_{\nu}$&$-i\frac{e^{2}(3C_{W}^{2}-2)}{C_{W}^{2}\sqrt{1-2S_{W}^{2}}}g_{\mu\nu}$
\\
\hline\hline
\end{tabular}
\end {center}
\begin{center}
 \null\noindent ~~{\bf Table 2:} Feynman rules for
 $\phi^{+}\phi^{-}V_{i}V_{j}$ vertices \cite{Hock}.
 \end {center}\vspace{0.1in}

\begin{center}
\doublerulesep 0.8pt \tabcolsep 0.1in
\begin{tabular}{||c|c|c||}\hline
\hline $i/j$ &vertices&$iE_{ij}^{\phi^{+}\phi^{-}}Q_{\mu}$\\
\hline $1/1$
&$\phi^{+}\phi^{-}A_{\mu}$&$-ie(p_{1}-p_{2})_{\mu}$\\
\hline $2/1$
&$\phi^{+}\phi^{-}Z_{\mu}$&$i\frac{eS_{W}}{C_{W}}(p_{1}-p_{2})_{\mu}$
\\ \hline $3/1$
&$\phi^{+}\phi^{-}Z_{H_{\mu}}$&$-i\frac{e(1-3S_{W}^{2})}{2S_{W}C_{W}\sqrt{1-2S_{W}^{2}}}(p_{1}-p_{2})_{\mu}$
\\
\hline\hline
\end{tabular}
\end {center}
 \null\noindent ~~{\bf Table 3:} Feynman rules for
 $\phi^{+}\phi^{-}V_{i}$ vertices. $p_{1}$ and $p_{2}$ refer to the out coming momentum of the first and second particle, respectively. \cite{Hock}.
\vspace{0.1in}

\begin{center}
\doublerulesep 0.8pt \tabcolsep 0.1in
\begin{tabular}{||c|c|c|c||}
\hline $hX_{1}X_{2}$&$V_{hX_{1}X_{2}}$&$X_{1}X_{2}h$&$V_{\phi^{+}\phi^{-}h}$\\
\hline $hZ_{\mu}Z_{\nu}$
&$eM_{W}g_{\mu\nu}/(C_{W}^{2}S_{W})$&$\phi^{-}\phi^{+}h$&$x(p_{3}\cdot
p_{3}+2p_{1}\cdot p_{2})/(3\sqrt{2}f)$\\
$hZ_{\mu}Z_{H_{\nu}}$
&$e^{2}fxg_{\mu\nu}/(\sqrt{2}C_{W}^{2}\sqrt{1-2S_{W}^{2}})$&&\\
\hline
\end{tabular}
\end {center}
 \null\noindent ~~{\bf Table 4:} Relevant coupling constants of the Higgs boson in Fig. 1(c). $p_{1}$, $p_{2}$ and $p_{3}$ refer to the incoming momentum of the first,second and third particle, respectively. \cite{Hock}.

\vspace{.5cm} \noindent{\bf Appendix B: The relevant coupling
constants in the process $e^{+}e^{-}\rightarrow Z\phi^{0}\phi^{0}$ }
 \vspace{0.1in}
\begin{center}
\doublerulesep 0.8pt \tabcolsep 0.1in
\begin{tabular}{||c|c|c||}\hline
\hline $i/j$ &vertices&$iC_{ij}^{\phi^{0}\phi^{0}}g_{\mu\nu}$\\
\hline $2/1$
&$\phi^{0}\phi^{0}Z_{\mu}Z_{\nu}$&$-i\frac{e^{2}x^{2}}{54C_{W}^{2}S_{W}^{2}}g_{\mu\nu}$\\
\hline $3/1$
&$\phi^{0}\phi^{0}Z_{H_{\mu}}Z_{\nu}$&$-i\frac{e^{2}x^{2}}{54C_{W}^{2}\sqrt{1-2S_{W}^{2}}}g_{\mu\nu}$\\
\hline\hline
\end{tabular}
\end {center}
\begin{center}
 \null\noindent ~~{\bf Table 5:} Feynman rules for
 $\phi^{0}\phi^{0}Z_{i}Z$ vertices \cite{Hock}.
 \end {center}\vspace{0.1in}
 \begin{center}
\doublerulesep 0.8pt \tabcolsep 0.1in
\begin{tabular}{||c|c|c|c||}
\hline $h\phi^{0}Z_{i}$&$V_{h\phi^{0}Z_{i}}$&$hX_{1}X_{2}$&$V_{h\phi^{0}\phi^{0}}$\\
\hline $h\phi^{0}Z_{\mu}$
&$iexp_{3_{\mu}}/6S_{W}C_{W}$&$h\phi^{0}\phi^{0}$&$\frac{x}{27\sqrt{2}f}(30p_{2}\cdot p_{3}+11p_{1}\cdot p_{1})$\\
$h\phi^{0}Z_{H_{\mu}}$
&$i\frac{ex}{18S_{W}C_{W}\sqrt{1-2S_{W}^{2}}}[(14-17S_{W}^{2})p_{2_{\mu}}-(4-S_{W}^{2})p_{1_{\mu}}]$&&\\
\hline
\end{tabular}
\end {center}
 \null\noindent ~~{\bf Table 6:} Relevant coupling constants of the neutral scalar.
 $p_{1}$, $p_{2}$ and $p_{3}$ refer to the incoming momentum of the first,second and third particle, respectively. \cite{Hock}.


\begin{thebibliography}{90}
\bibitem{hierarchy-1}
H. Georgi and A. Pais, {\it Phys. Rev.} D{\bf 10} (1974) 539; {\it
Phys. Rev.} D{\bf 12} (1975) 508.
\bibitem{hierarchy-2}
D. B. Kaplan and H. Georgi, {\it Phys. Lett.} B{\bf 136} (1984) 183;
D. B. Kaplan, H. Georgi and S. Dimopoulos, {\it Phys. Lett.} B{\bf
136} (1984) 187; H. Georgi and D. B. Kaplan, {\it Phys. Lett.} B{\bf
145} (1984) 216.
\bibitem{little-1}
N. Arkani-Hamed, A. G. Cohen, and H. Georgi, {\it Phys. Lett.} B{\bf
513} (2001) 232.
\bibitem{little-2}
N. Arkani-Hamed, A. G. Cohen, E. Katz, A. E. Nelson, T. Gregoire,
and J. G. Wacker, {\it JHEP} {\bf 0208} (2002) 021; I. Low, W.
Skiba, and D. Smith, {\it Phys. Rev.} D{\bf 66} (2002) 072001; D. E.
Kaplan and M. Schmaltz, {\it JHEP} {\bf 0310} (2003) 039.
\bibitem{ly-1}
Z. Chacko, H. S. Goh and R. Harnik, {\it Phys. Rev. Lett} {\bf 96}
 (2006) 231802; Z. Chacko, Y. Nomura, M. Papucci and G. Perez, {\it
JHEP} {\bf 0601} (2006) 126.
\bibitem{ly-2}
Z. Chacko, H. S. Goh and R. Harnik, {\it JHEP} {\bf 0601} (2006)
108.
\bibitem{ly-3}
A. falkowski, S. Pokorski and M. Schmaltz, {\it Phys. Rev.} D{\bf
74} (2006) 035003.
\bibitem{Hock}
H.S. Goh and S. Su,  {\it Phys. Rev.} D{\bf 75} (2007) 075010.
\bibitem{dong}
Hock-Seng Goh and C. A. Krenke, {\it Phys. Rev.} D {\bf 76} (2007)
115018; {\it Phys. Rev.} D{\bf 81} (2010) 055008; A. Abada and I.
Hidalgo, {\it Phys. Rev.} D {\bf 77} (2008) 113013; E. M. Dolle and
Shufang Su, {\it Phys. Rev.} D{\bf 77} (2008) 075013; Yao-Bei Liu,
Xue-Lei Wang, Hong-Mei Han and Yong-Hua Cao, {\it Commun. Theor.
Phys.} {\bf 49}(2008) 977; Yao-Bei Liu and Jie-Fen Shen, {\it Mod.
Phys. Lett.} A{\bf 24} (2009) 143; Chong-Xing Yue, Hui-Di Yang and
Wei Ma, {\it Nucl. Phys.} B{\bf 818} (2009) 1; P. Batra and Z.
Chacko, {\it Phys. Rev.} D{\bf 79} (2009) 095012; Hock-Seng Goh, C.
A. Krenke, {\it Phys. Rev.} D{\bf 81} (2010) 055008; Lei Wang and
Jin Min Yang, {\it JHEP} {\bf 1005} (2010) 024.
\bibitem{loop}
Dong-Won Jung and Jae-Young Lee, hep-ph/{\bf 0701071}.
\bibitem{liu}
Yao-Bei Liu, Xue-Lei Wang, Jun Cao and Hong-Mei Han, {\it Commun.
Theor. Phys.} {\bf 50}(2008) 445; Yao-Bei Liu, Lin-Lin Du and Qin
Chang, {\it Mod. Phys. Lett.} A{\bf 24} (2009) 463; Yao-Bei Liu,
Shuai-Wei Wang, {\it Int. J. Mod. Phys.} A{\bf 24} (2009) 4261;
Yao-Bei Liu and Xue-Lei Wang, {\it Europhys. Lett.} {\bf 86}, 61002
(2009).
\bibitem{liu1}
Yao-Bei Liu, Hong-Mei Han and Xue-Lei Wang, {\it Eur. Phys. J. C}
{\bf 53}(2008) 615.
\bibitem{yue} Wei Ma, Chong-Xing Yue and Yong-Zhi Wang, {\it
Phys. Rev.} D {\bf 79} (2009) 095010.
\bibitem{ILC}
J. Brau (Ed.) et al, By ILC Collaboration, {\it LC Reference Design
Report: ILC Global Design Effort and World Wide Study.,}
FERMILA-APC, Aug 2007, arXiv: acc-ph/{\bf 0712.1950}.
\bibitem{CLIC}
CLIC Physics Working Group (E. Accomando et al.), hep-ph/{\bf
0412251}.
\bibitem{sm}
J. J. Lopez-Villarejo, J. A. M. Vermaseren, arXiv:
0812.3750[hep-ph]; A. Djouadi, V. Driesen, C. Junger, {\it Phys.
Rev.} D {\bf 54} (1996) 759.
\bibitem{bsm}
A. Djouadi, H. E. Haber, P. M. Zerwas, {\it Phys. Lett.} B{\bf 375}
(1996) 203; J. L. Feng, T. Moroi, {\it Phys. Rev.} D {\bf 56} (1997)
5962; H. Grosse, Yi Liao, {\it Phys. Rev.} D {\bf 64} (2001) 115007;
N. Delerue, K. Fujii, N. Okada, {\it Phys. Rev.} D {\bf 70} (2004)
091701; Yao-Bei Liu, Lin-Lin Du and Xue-Lei Wang, {\it J. Phys.} G
{\bf 33} (2007) 577;  A. Arhrib, R. Benbrik, C. W. Chiang, {\it
Phys. Rev.} D {\bf 77} (2008) 115013; Yao-Bei Liu, Xue-Lei Wang and
Hong-Mei Han, {\it Europhys. Lett.} {\bf 81} (2008) 31001; R. N.
Hodgkinson, D. Lopez-Val, Joan Sola, {\it Phys. Lett.} B{\bf 673}
(2009) 47; A. Cagil and M. T. Zeyrek,{\it Phys. Rev.} D {\bf 80}
(2009) 055021; A. Gutierrez-Rodriguez, M. A. Hernandez-Ruiz, O. A.
Sampayo, arXiv:{\bf 0903.1383} [hep-ph].
\bibitem {HZ}
K. Hagiwara, D. Zeppenfeld. {\it Nucl. Phys.} B{\bf 313}, (1989)560;
V. Barger, Han Tao, D. Zeppenfeld. {\it Phys. Rev.} D{\bf 41},
(1990)2782.
\bibitem{calchep}
A. Pukhov et al., hep-ph/{\bf9908288}; hep-ph/{\bf0412191}.
\bibitem{data}
C. Amsler et al. [Particle Data Group] {\it Phys. Lett.} B{\bf 667}
(2008) 1.
\bibitem{adj}
A. Djouadi, et al., {\it Eur. Phys. J.} C {\bf 10} (1999)27; E.
Coniavitis, A. Ferrari, {\it Phys. Rev.} D {\bf 75} (2007) 015004;
U. Baur, {\it Phys. Rev.} D{\bf 80} (2009) 013012; Y. Takubo,
arXiv:{\bf 0901.3598} [hep-ph]; arXiv:{\bf 0907.0524} [hep-ph].
\end{thebibliography}
\end{document}